\documentclass[9pt]{book}

\usepackage[dvips]{graphicx,color} 
\usepackage{makeidx,far,detector}
\usepackage{amssymb,amsmath,epsfig,epsf,subfigure}

\makeauthorindex

\BookTitle{Far Detector in Korea for the J-PARC Neutrino Beam}
\CopyRight{\copyright 2007 by Universal Academy Press, Inc.}

\begin{document}
\pagenumbering{arabic}

\chapter{
Probing Nonstandard Neutrino Physics at T2KK }

\author{\raggedright \baselineskip=10pt%
{\bf
N. Cipriano Ribeiro$^1$, T. Kajita$^2$, 
P. Ko$^3$, H. Minakata$^4$,
S. Nakayama$^5$, and H. Nunokawa$^1$ }\\ 
{\small \it %
(1)  Departamento de F\'{\i}sica, Pontif{\'\i}cia Universidade Cat{\'o}lica
do Rio de Janeiro, C. P. 38071, 22452-970, Rio de Janeiro, Brazil\\
(2) Research Center for Cosmic Neutrinos, Institute for Cosmic Ray Research,
and Institute for the Physics and Mathematics of the Universe, University of Tokyo, Kashiwa, Chiba 277-8582, Japan \\
(3) School of Physics, KIAS, Seoul 130-722, Korea
\\
(4) Department of Physics, Tokyo Metropolitan University, Hachioji,
Tokyo 192-0397, Japan
\\
(5)  Kamioka Observatory, Institute for Cosmic Ray Research, Univ. of Tokyo,
Higashi-Mozumi, Kamioka-cho, Hida, Gifu 506-1205, Japan}
}


\AuthorContents{N.\ Cipriano, T.\ Kajita, P.\ Ko, H.\ Minakata,
S.\ Nakayama and H.\ Nunokawa} 

\AuthorIndex{Cipriano Ribeiro}{N.} 

\AuthorIndex{Kajita}{T.} 

\AuthorIndex{Ko}{P.} 

\AuthorIndex{Minakata}{H.}

\AuthorIndex{Nakayama}{S.}

\AuthorIndex{Nunokawa}{H.}

     \baselineskip=10pt
     \parindent=10pt

\section*{Abstract}

Having a far detector in Korea for the J-PARC neutrino beam in
addition to one at Kamioka has been shown to be a
powerful way to lift neutrino parameter ($\Delta m^2$ and mixing
angles) degeneracies. In this talk, I report the sensitivity of
the same experimental setup to nonstandard neutrino physics, such
as quantum decoherence, violation of Lorentz symmetry
(with/without CPT invariance), and nonstandard neutrino
interactions with matter. In many cases, two detector setup is
better than one detector setup at SK. This observation makes
another support for the two detector setup.

\section{Introduction}
\label{introduction}

The neutrino mass induced neutrino oscillation has been identified
as a dominant mechanism for neutrino disappearances through a
number of the neutrino experiments: the atmospheric \cite{SKatm},
solar \cite{solar}, reactor \cite{KamLAND}, and accelerator
\cite{accelerator} experiments. After passing through the
discovery era, the neutrino physics will enter the epoch of
precision study, as the CKM phenomenology and CP violation in the
quark sector. The MNS matrix elements will be measured with higher
accuracy, including the CP phase(s), and the neutrino properties
such as their interactions with matter etc. will be studied in a
greater accuracy.

In recent years \cite{T2KK1st,T2KK2nd}, the physics potential of
the Kamioka-Korea two detector setting which receive an intense
neutrino beam from ~{J-PARC} was considered in detail, and it has
been demonstrated that the two detector setting is powerful enough
to resolve all the eight-fold parameter degeneracy
\cite{intrinsic,MNjhep01,octant}, if $\theta_{13}$ is in reach of
the next generation accelerator \cite{T2K,NOVA} and the reactor
experiments \cite{MSYIS,reactor13}. The degeneracy includes the
parameters $\theta_{13}$, $\delta$ and octant of $\theta_{23}$,
and it is doubled by the ambiguity which arises due to the unknown
sign of $\Delta m^2_{31}$. The detector in Korea plays a decisive
role to lift the last one. For related works on Kamioka-Korea two
detector complex, see, for example,
\cite{hagiwara,Okumura-Seoul2006,Dufour-Seoul2006,Rubbia-Seoul2006}.
This observation was the main motivation for this series of
workshops, and a lot of speakers gave talks about physics
potential at the Kamioka-Korea two detector setup in the past
three workshops.

During the course of precision studies, it will become natural to
investigate nonstandard physics related with neutrinos. In this
talk, I will show that the Kamioka-Korea identical two detector
setting is also a unique apparatus for studying nonstandard
physics (NSP), by demonstrating that the deviation from the
expectation by the standard mass-induced oscillation can be
sensitively probed by comparing yields at the intermediate
(Kamioka) and the far (Korea) detectors. In this talk, I discuss
the potential of the Kamioka-Korea setting, concentrating on
$\nu_\mu - \nu_\tau$ subsystem in the standard three-flavor mixing
scheme, and focus on $\nu_\mu$ disappearance measurement. We
consider three different types of nonstandard neutrino physics:
\begin{itemize}
\item Quantum decoherence (QD) \cite{ellis,QD-bari1,benatti-floreanini}

\item Tiny violation of Lorentz symmetry with/without CPT
\cite{Coleman1,Coleman2,Kostelecky}

\item Nonstandard neutrino interactions of neutrinos with matter
due to some new physics \cite{wolfenstein,grossmann}
\end{itemize}
The first two cases go beyond the conventional quantum field
theory framework, whereas the last case is strictly within the
conventional QFT.

In analyzing the nonstandard physics, we aim at demonstrating the
powerfulness of the Kamioka-Korea identical two detector setting,
compared to other settings. For this purpose, we systematically
compare the results obtained with the following three settings
(the number indicate the fiducial mass):

\begin{itemize}

\item

Kamioka-Korea setting:
Two identical detectors one at Kamioka and the other in Korea each 0.27 Mton

\item
Kamioka-only setting: A single 0.54 Mton detector at Kamioka

\item
Korea-only setting: A single 0.54 Mton detector at somewhere in Korea.

\end{itemize}
Among the cases we have examined Kamioka-Korea setting always
gives the best sensitivities, apart from two exceptions of violation
of Lorentz invariance in a CPT violating manner, and the nonstandard
neutrino interactions with matter.
Whereas, the next best case is sometimes Kamioka-only or Korea-only settings
depending upon the problem.

This talk is organized as follows. In Sec.~\ref{NSP}, we
illustrate how we can probe nonstandard physics with Kamioka-Korea
two detector setting, with a quantum decoherence as an example of
nonstandard neutrino physics.
In Sec.~\ref{decoherence}, we discuss quantum decoherence.
In Sec.~\ref{lorentz}, we discuss possible violation of Lorentz invariance.
In Sec.~\ref{NSI}, we discuss non-standard neutrino matter interactions,
and the results of study is summarized in Sec.~\ref{conclusion}.
This talk is based on the work \cite{ckkmnn}, where one can find
more plots and detailed discussions covered in this talk.

\section{Basic ideas \label{NSP}}

\begin{figure}[bhtp]
\begin{center}
\includegraphics[height=10.0cm]{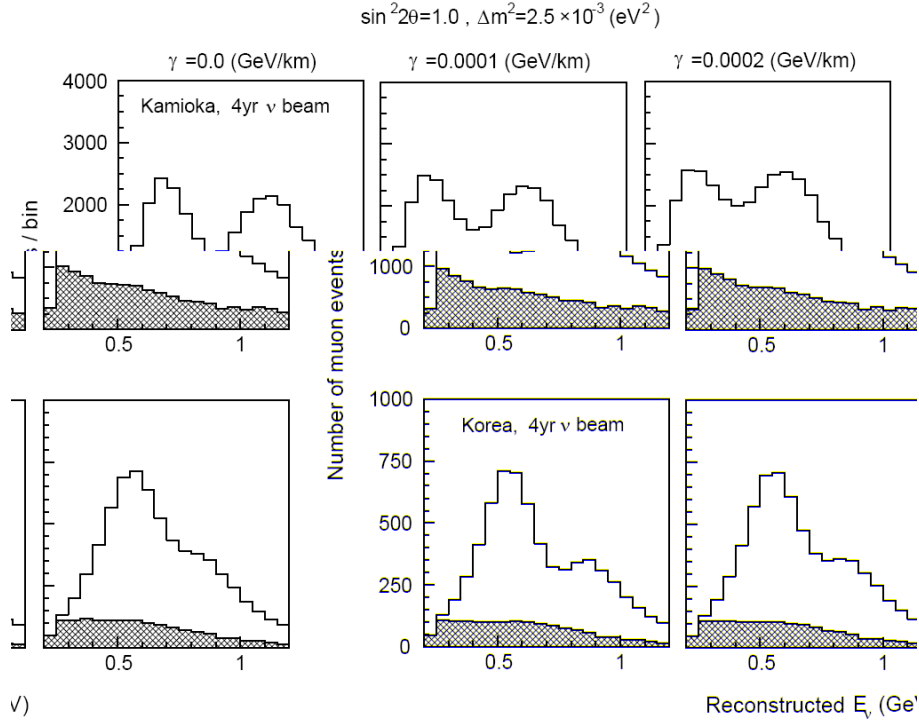}
\end{center}
\caption{\label{example_event}
Event spectra of neutrinos at Kamioka (the top panel) and Korea
(the bottom panel)
for $\gamma = 0$ (the left column), $1 \times 10^{-4}$ GeV/km
(the middle column), and $\gamma = 2 \times 10^{-4}$ GeV/km
(the right column). The hatched areas denote the non-quasi-elastic events.}
\end{figure}

Let me first describe the basic strategy of our analysis adopted
in the following sections. For the purpose of illustration, we
consider quantum decoherence (QD), for which the $\nu_\mu$
survival probability is given by
\begin{equation}
P ( \nu_\mu \rightarrow \nu_\mu ) = P ( \overline{\nu}_\mu
\rightarrow \overline{\nu}_\mu ) = 1 - {1\over 2}~\sin^2 2 \theta
~ \left[ 1 - e^{-\gamma (E) L} \cos \left( {\Delta m^2 L \over 2
E} \right) \right],
\end{equation}
with $\gamma(E) = \gamma / E$ as an illustration. The
$\overline{\nu}_\mu$ survival probability is the same as above,
assuming CPT invariance in the presence of QD. Then one can
calculate the number of $\nu_\mu$ and $\overline{\nu}_\mu$ events
observed at two detectors placed at Kamioka and Korea, using the
above survival probability and the neutrino beam profiles. For
simplicity, let us consider the number of observed neutrino events
both at Kamioka and Korea, for each energy bin (with 50 MeV width)
from $E_\nu = 0.2$ GeV upto $E_\nu = 1.4$ GeV. In
Fig.~\ref{example_event}, we show the $\nu_\mu$ event spectra at
detectors located at Kamioka and Korea for the pure oscillation
$\gamma = 0$ (the left column) and the oscillation plus QD with
two different QD parameters, $\gamma = 1 \times 10^{-4}$ GeV/km
(the middle column) and $\gamma = 2 \times 10^{-4}$ GeV/km (the
right column). \footnote{In order to convert this $\gamma$ in unit
of GeV/km to $\gamma$ defined in Eq.~(2), one has to multiply
$0.197 \times 10^{-18}$.} One observes the spectral distortion for
non-vanishing $\gamma$. In particular, the spectral distortions
are different between detectors at Kamioka and Korea due to the
differect $L/E$ values at the two positions.

Assuming the actual data at Kamioka and Korea are given (or well
described) by the pure oscillation with $\sin^2 2 \theta = 1$ and
$\Delta m^2 = 2.5 \times 10^{-2}$ eV$^2$, we could claim that
$\gamma = 1 \times 10^{-4}$ GeV/km (shown in the middle column),
for example, would be inconsistent with the data. One can make
this kind of claim in a more proper and quantitative manner using
the $\chi^2$ analysis, which is described in details in
Ref.~\cite{ckkmnn}.

\section{Quantum Decoherence (QD) \label{decoherence}}

\subsection{Motivation}

When a quantum system interacts with environment, quantum
decoherence (QD) could appear. A classic example is a two-slit
experiment with electron beams. If we do not measure which hole an
electron passes through, one observes an interference pattern. On
the other hand, if we try to determine which hole an electron
passes through using some device, the interference pattern will be
distorted. As the disturbance becomes stronger, the interference
will be distorted more, thus it eventually disappears.

It has been speculated for some time that there may be a loss of
quantum coherence due to environmental effect or quantum gravity
and space-time foam, etc.. Although quantum decoherence (QD) due
to rapid fluctuation of environment is conceivable, QD due to
quantum gravity is still under debate among theoreticians. In this
talk, I present our phenomenological study of QD, namely how this
effect can be probed by the Kamioka-Korea setting. rather than
discuss the ground on the origin  of QD within quantum gravity.
For previous analyses of decoherence in neutrino experiments, see
e.g., \cite{QD-bari1,QD-bari2,barenboim}.

As discussed in Sec.~\ref{introduction} we consider the
$\nu_\mu - \nu_\tau$ two-flavor system.
Since the matter effect is a sub-leading effect in this channel
we employ vacuum oscillation approximation in this section.
The two-level system in vacuum in the presence of quantum
decoherence can be solved to give the $\nu_{\mu}$ survival
probability Eq. (1) \cite{QD-bari1,benatti-floreanini}.
Notice that the conventional two-flavor oscillation formula
is reproduced in the limit $\gamma (E) \rightarrow 0$.
Since the total probability is still conserved in the presence of QD,
the relation
$P(\nu_{\mu} \rightarrow \nu_{\tau}) = 1- P(\nu_{\mu} \rightarrow \nu_{\mu})$
holds.

Nothing is known for the energy dependence of $\gamma (E)$ from
the first principle including quantum gravity. Therefore, we
examine, following \cite{QD-bari1}, several typical cases of
energy dependence of $\gamma (E)$, which are purely
phenomenological ansatzs:
\begin{eqnarray}
\gamma (E) = \gamma
\left( \frac{ E }{ \text{ GeV } } \right)^n ~({\rm with}~ n=0,2,-1)
\label{E-dep}
\end{eqnarray}
In this convention, the overall constant $\gamma$ has a dimension of
energy or (length)$^{-1}$, for any  values of the exponent $n$.
We will use $\gamma$ in GeV unit in this section.
In the following three subsections, we analyze three different
energy dependences, $n=0, -1, 2$ one by one.

\subsection{Numerical Results and Discussions}

\begin{figure}[bhtp]
\begin{center}
\includegraphics[width=0.40\textwidth]{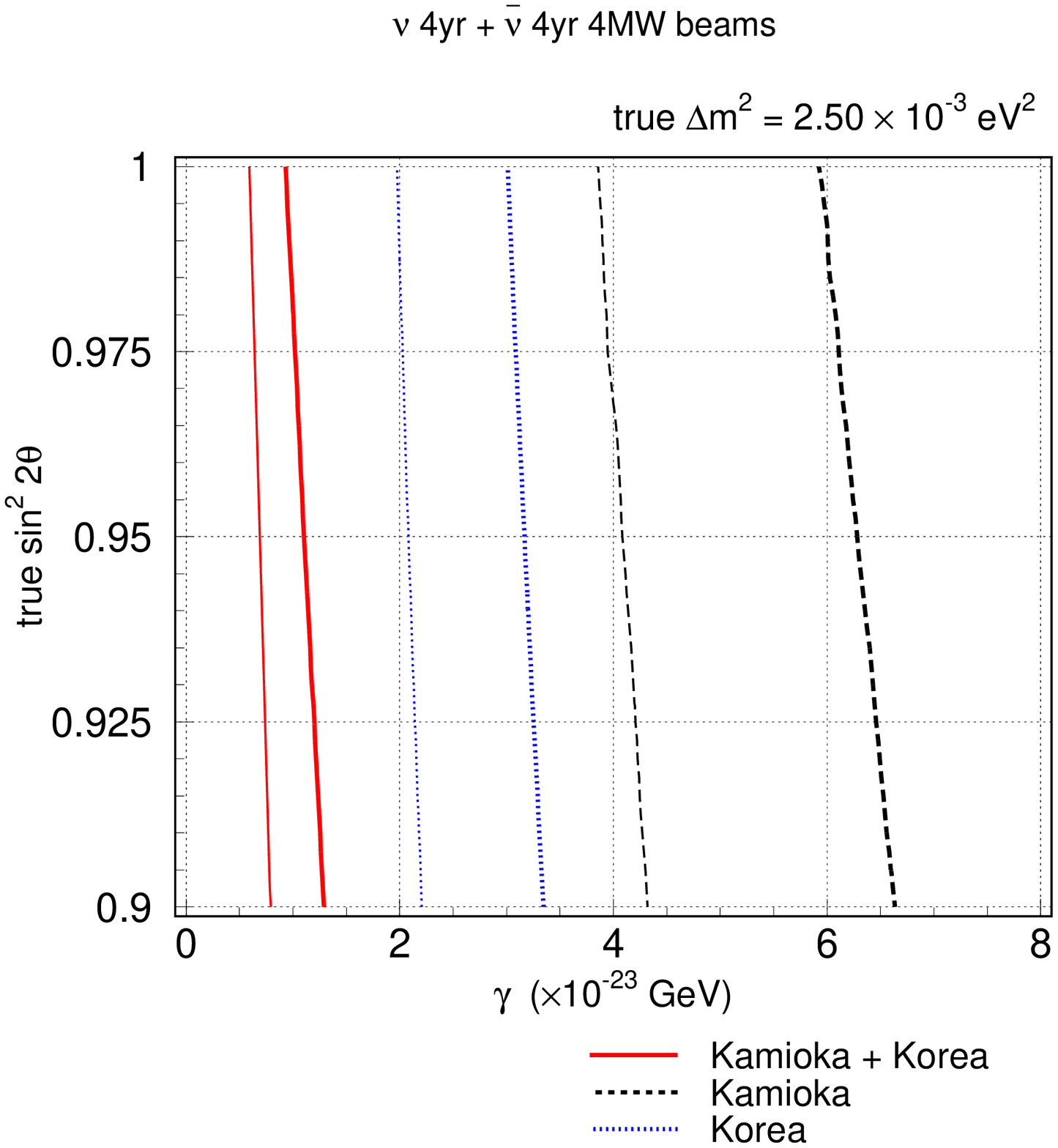}
\includegraphics[width=0.40\textwidth]{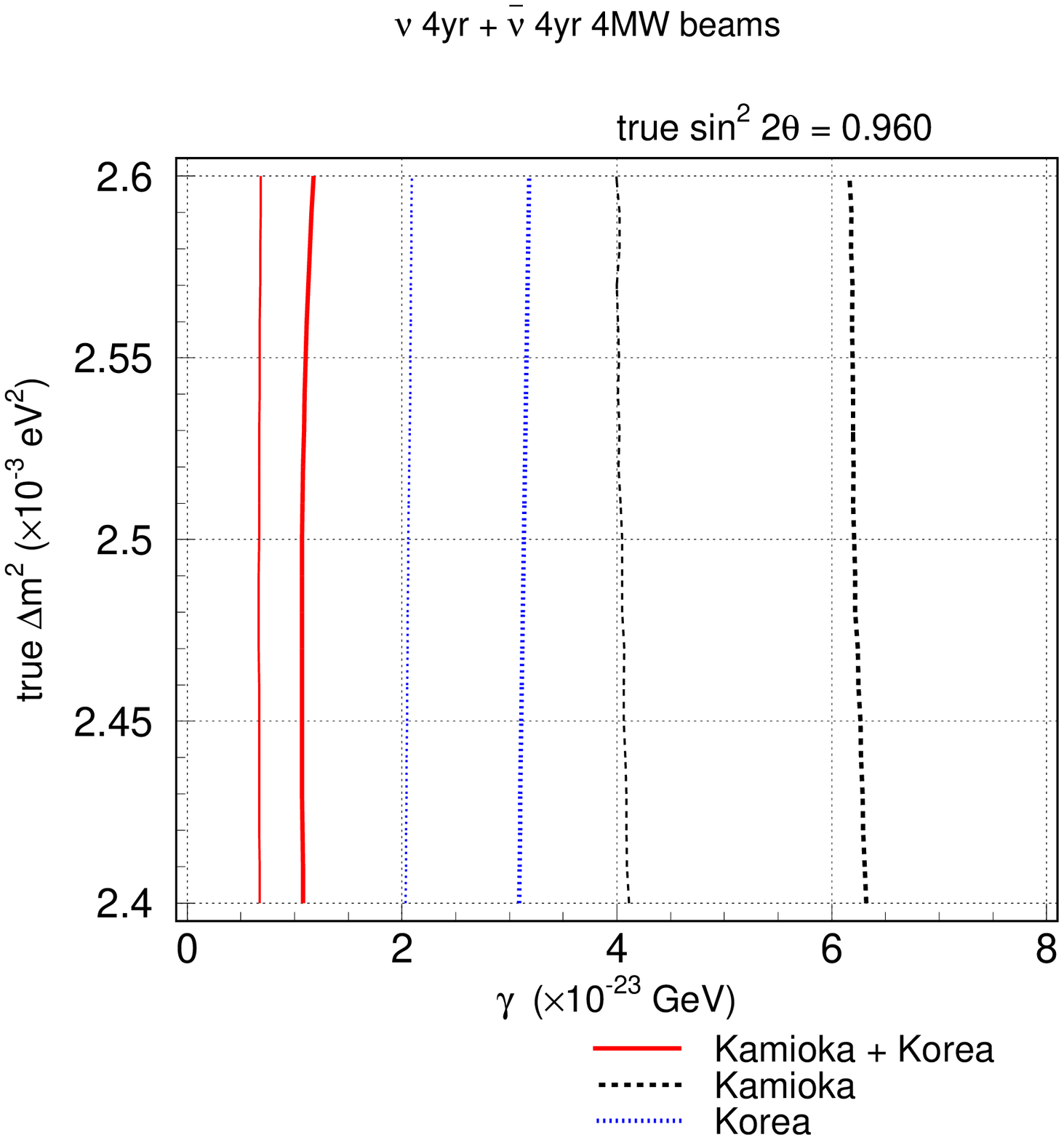}
\end{center}
\caption{The sensitivity to $\gamma$ as a function of
$\sin^2 2\theta \equiv \sin^2 \theta_{23}$ (left panel) and
$\Delta m^2 \equiv \Delta m^2_{32}$ (right panel).
The case of 1/E dependence of $\gamma(E)$.
The red solid lines are for Kamioka-Korea setting with each 0.27
Mton detector, while the dashed black (dotted blue) lines are for
Kamioka (Korea) only setting with 0.54 Mton detector. The thick
and the thin lines are for 99 \% and 90 \% CL, respectively. 4
years of neutrino plus 4 years of anti-neutrino running are
assumed.
The other input values of the parameters are
$\Delta m_{31}^2 = +2.5\times 10^{-3}$ eV$^2$
(with positive sign indicating the normal mass hierarchy)
and $\sin ^2 \theta_{23}$=0.5.
The solar mixing parameters are fixed as
$\Delta m_{21}^2 = 8\times 10^{-5}$ eV$^2$ and $\sin^2{\theta_{12}}$=0.31.
}
\label{decoh-gamma-1-over-E}
\end{figure}

First, let me consider the case with $n=-1$: $\gamma (E) \propto
\frac{1}{E}$. 
It turns out that the correlations between $\Delta m^2$ and
$\sin^2 2 \theta$ at three experimental setups. Note that there
are strong correlations between $\sin^2 2 \theta$ and $\gamma$ for
the Kamioka-only and Korea-only setups, and the slope of the
correlation for the Kamioka-only setup is different from that for
the Korea-only setup (see Fig.~2 in Ref.~\cite{ckkmnn}). Therefore
the Kamioka-Korea setup can give a stronger bound than each
experimental setup. This advantage can be seen in
Fig.~\ref{decoh-gamma-1-over-E}, where we present the sensitivity
regions of $\gamma$ as a function of $\sin^2 2\theta$ (left 
panel) and $\Delta m^2$ (right panel).

We can repeat the same analysis for other cases $n=0$ and $n= 2$.
In Table \ref{table-gamma}, we summarize the bounds on $\gamma$ at
2$\sigma$ CL achievable by three different experimental settings,
along with the upper bounds on $\gamma$ at 90\% CL obtained by
analyzing the atmospheric neutrino data in \cite{QD-bari1}, for
the purpose of comparison.

In the case of $\frac{1}{E}$ dependence of $\gamma (E)$, the
sensitivity to $\gamma$ in the Kamioka-Korea setting is better
than the Korea-only and the Kamioka-only settings by a factor
greater than 3 and 6, respectively. Also all the three settings
can improve the current bound almost by two orders of magnitude.
This case demonstrates clearly that the two-detector setup is more
powerful than the Kamioka-only setup.

For $n=0$ (the case of an energy independent constant $\gamma
(E)$), we find that the sensitivity to $\gamma$ in the
Kamioka-Korea setting is better than the Korea-only and the
Kamioka-only settings by a factor greater than 3 and 8,
respectively. Also Kamioka-Korea two detector setting can improve
the current bound by a factor of $\sim 3$.

Finally, for $n=2$ with $\gamma (E) \propto E^2$, the sensitivity
to $\gamma$ in the Kamioka-Korea setting is better than the
Korea-only and the Kamioka-only settings by a factor greater than
3 and 5, respectively. On the other hand, the current bound
imposed by the atmospheric neutrino data surpasses those of our
three settings by almost $\sim$4 orders of magnitude. Because the
spectrum of atmospheric neutrinos spans a wide range of energy
which extends to 100-1000 GeV, it gives much tighter constraints
on the decoherence parameter for quadratic energy dependence of
$\gamma (E)$. In a sense, the current Super-Kamiokande experiment
is already a powerful neutrino spectroscope with a very wide
energy range, and could be sensitive to nonstandard neutrino
physics that may affect higher energy neutrinos such as QD with
$\gamma (E) \sim E^2$.

\begin{table}[t]
\small \caption{\label{table-gamma} Presented are the upper bounds
on decoherence parameters $\gamma$ defined in (\ref{E-dep}) for
three possible values of $n$. The current bounds are based on
\cite{QD-bari1} and are at 90\% CL. The sensitivities obtained by
this study are also at 90 \% CL , and correspond to the true
values of the parameters $\Delta m^2=2.5 \times 10^{-3} {\rm
eV}^2$ and $\sin^2 2\theta_{23} = 0.96$.}
\begin{center}
\begin{tabular}{|c|c|c|c|c|}
\hline
$n$ & Curent bound &
Kamioka-only & Korea-only & Kamioka-Korea
\\
\hline
$n = 0$ & $< 3.5 \times 10^{-23}$  &
$< 8.7 \times 10^{-23}$   & $< 3.2 \times 10^{-23}$   &
$< 1.1 \times 10^{-23}$
\\
$n = -1$ & $< 2.0 \times 10^{-21}$  &
$< 4.0 \times 10^{-23} $ & $< 2.0 \times 10^{-23}$  & $ < 0.7 \times 10^{-23}$
\\
$n = 2 $  & $< 0.9 \times 10^{-27}$  &
$< 9.2 \times 10^{-23}$ & $< 6.0 \times 10^{-23}$ & $< 1.7 \times 10^{-23}$
\\
\hline
\end{tabular}
\end{center}
\end{table}

\section{Violation of Lorentz Symmetry \label{lorentz}}

\subsection{Motivation}

Lorentz symmetry is one of the cornerstones of the quantum field
theory, which is a mathematical tool for high energy physics
nowadays. Therefore it is important to test this symmetry
experimentally as accurately as possible. There may be a small
violation of Lorentz symmetry, which would modify the usual
energy-momentum dispersion relations. In such a case, neutrinos
can have both velocity mixings and the mass mixings, which are CPT
conserving \cite{Coleman1}. Also there could be CPT-violating
interactions in general \cite{Coleman1,Coleman2,Kostelecky}. Then,
the energy of neutrinos with definite momentum in
ultra-relativistic regime can be written as
\begin{eqnarray}
\frac{m m^{\dagger}} {2p}  = c p + {m^2 \over 2 p} + b,
\label{rel-energy}
\end{eqnarray}
where $m^2$, $c$, and $b$ are $3 \times 3$ hermitian matrices, and the
three terms represent, in order, the effects of velocity mixing,
mass mixing, and CPT violation \cite{Coleman2}.
The energies of neutrinos are eigenvalues of (\ref{rel-energy}),
and the eigenvectors give the ``mass eigenstates''.
Notice that while $c$ is dimensionless quantity, $b$ has
dimension of energy.
For brevity, we will use GeV unit for $b$.

Within the framework just defined above, we can work in the
$\nu_\mu - \nu_\tau$ two flavor subsystem, and derive the
$\nu_\mu$ survival probability, which depends on six parameters.
We further assume that three matrices $m^2$, $c$ and $b$ are
diagonalized by the same unitarity transformations with the same
mixing angles: namely, $\theta_m = \theta_c = \theta_b \equiv
\theta$. Then the $\nu_\mu$ survival probability is given by :
\begin{eqnarray}
P ( \nu_\mu \rightarrow \nu_\mu ) = 1 - \sin^2 2 \theta~ \sin^2
\left[ L \left( {\Delta m^2 \over 4 E} + {\delta b \over 2} +
{\delta c E \over 2} \right) \right], \label{FLY}
\end{eqnarray}
and we recovers the case treated in \cite{Foot:1998vr}.
Here, $\delta b \equiv b_2 - b_1$ and $\delta c \equiv c_2 - c_1$,
where $c_{i=1,2}$ and $b_{i=1,2}$, are the eigenvalues of the
matrix $c$ and $b$. Note that we still have 4 parameters,
$\theta$, $\Delta m^2$, $\delta b$ and $\delta c$.
The survival probability for the anti-neutrino is obtained
by the following substitution:
\begin{eqnarray}
\delta c \rightarrow \delta c , ~~~ \delta b \rightarrow - \delta b
\end{eqnarray}
The difference in the sign changes signify the CPT conserving vs. CPT
violating nature of $c$ and $b$ terms.
As pointed out in \cite{glashow}, the analysis for violation of
Lorentz invariance with $\delta c$ term is equivalent to testing
the equivalence principle \cite{equivalence}.
The oscillation probability in (\ref{FLY}) looks like the one for
conventional neutrino oscillations due to
$\Delta m^2$, with small corrections due to the Lorentz symmetry violating
$\delta b$ and $\delta c$ terms. In this sense, it may be the most
interesting case to examine as a typical example with the Lorentz symmetry
violation. Note that the sign of $\delta b$ and $\delta c$ can have
different effects on the survival probabilities, so that the bounds on
$\delta b$ and $\delta c$ could depend on their signs, although we will
find that the difference is rather small.

\subsection{Numerical Results and Discussions}

For ease of analysis and simplicity of presentation, we further
restrict our analysis to the case of either $\delta b=0$ and
$\delta c \neq 0$ (CPT conserving), or $\delta b \neq 0$ and
$\delta c=0$ (CPT violating).

Let me first examine violation of Lorentz invariance with CPT
conservation, namely $\delta b=0$ and $\delta c \neq 0$. Unlike
the case of quantum decoherence, the sensitivities to $\delta c$
achieved by the Kamioka-Korea setting is slightly better than
those of the Korea-only and the Kamioka-only settings but not so
much. The sensitivity is weakly correlated to $\theta$, and the
best sensitivity is achieved at the maximal $\theta$. There is
almost no correlation to $\Delta m^2$.

Next we consider the CPT and Lorentz violating case ($\delta c=0$
and $\delta b \neq 0$). In this case, unlike the system with
decoherence, the sensitivity is greatest in the Kamioka-only
setting, though the one by the Kamioka-Korea setting is only
slightly less by about $15-20$\%. Whereas, the sensitivity by the
Korea-only setting is much worse, more than a factor of 2 compared
to the Kamioka-only setting. The reason for this lies in the
$\nu_\mu$ and $\overline{\nu}_\mu$ survival probabilities. In this
scenario, the effect of the nonvanishing $\delta b$ appears as the
difference in the oscillation frequency between neutrinos and
anti-neutrinos, if the energy dependence is neglected. In this
case, the measurement at different baseline is not very important.
Then the Kamioka-only setup turns out to be
slightly better than the Kamioka-Korea setup.
This case is also unique by having the worst sensitivity at the
largest value of $\Delta m^2$.
Also, the correlation of sensitivity to $\sin^2 2\theta$ 
is strongest among the cases examined in this paper,
with maximal sensitivity at maximal $\theta$. (See the right and 
the left panels of Fig.~5 of Ref.~\cite{ckkmnn} for details.)

I summarize the results in Table~2, along with the present bounds
on $\delta c$ and $\delta b$, respectively. We quote the current
bounds on $\delta c$'s from Ref.s~\cite{Fogli:1999fs,macro_LV}
which was obtained by the atmospheric neutrino data,
\begin{equation}
|\delta c_{\mu \tau}| \lesssim 3 \times 10^{-26}.
\end{equation}
We note that the current bound on $\delta c_{\mu \tau}$ obtained
by atmospheric neutrino data is quite strong. The reason why the
atmospheric neutrino data give much stronger limit is that the
relevant energy is much higher (typically $\sim 100$ GeV) than the
one we are considering ($\sim$ 1 GeV) and the baseline is larger,
as large as the Earth diameter.

For the bound on $\delta b$,  Barger et al. \cite{barger} argue
that
\begin{equation}
|\delta b_{\mu\tau}| < 3 \times 10^{-20}~{\rm GeV}
\end{equation}
from the analysis of the atmospheric neutrino data.

Let me compare the sensitivity on $\delta b$ within our two
detector setup with the sensitivity at a neutrino factory.  Barger
et al. \cite{barger} considered a neutrino factory with $10^{19}$
stored muons with 20 GeV energy, and 10 kton detector, and
concluded that it can probe $\delta b < 3 \times 10^{-23}$ GeV.
The Kamioka-Korea two detector setup and Kamioka-only setup have
five and six times better sensitivities compared with the neutrino
factory with the assumed configuration. Of course the sensitivity
of a neutrino factories could be improved with a larger number of
stored muons and a larger detector. A more meaningful comparison
would be possible, only when one has configurations for both
experiments which are optimized for the purposes of each
experiment. Still we can conclude that the Kamioka-Korea
two-detector setup could be powerful to probe the Lorentz symmetry
violation.

\begin{table}[t]
\small \caption{\label{table-liv} Presented are the upper bounds
on the velocity mixing parameter $\delta c$ and the CPT violating
parameter $\delta b$ (in GeV). The current bounds are based on
\cite{Fogli:1999fs,macro_LV,barger} and are at 90\% CL. The
sensitivites obtained in this study are also at 90 \% CL , and
correspond to the true values of the parameters $\Delta m^2=2.5
\times 10^{-3} {\rm eV}^2$ and $\sin^2 2\theta_{23} = 0.96$. }
\begin{center}
\begin{tabular}{|c|c|c|c|c|}
\hline
LV parameters & Curent bound &
Kamioka-only  & Korea-only &  Kamioka-Korea
\\
\hline $| \delta c | $ & $< 3 \times 10^{-26}$  & $\lesssim 5
\times 10^{-23}$   & $\lesssim 4 \times 10^{-23}$   & $\lesssim 3
\times 10^{-23}$
\\
$| \delta b |$ (GeV) & $ < 3.0 \times 10^{-20}$  &
$\lesssim 1 \times 10^{-23} $ & $\lesssim 0.5 \times 10^{-23}$  &
$\lesssim 0.6 \times 10^{-23}$
\\
\hline
\end{tabular}
\end{center}
\end{table}

\section{Nonstandard Neutrino Interactions with Matter \label{NSI}}
\subsection{Motivations}

In the presence of new physics around electroweak scale, neutrinos
might have nonstandard neutral current interactions with matter
\cite{wolfenstein,grossmann,berezhiani,NSI}, $\nu_\alpha + f
\rightarrow \nu_\beta + f$ ($\alpha, \beta = e, \mu , \tau$), with
$f$ being the up quarks, the down quarks and electrons.
In such a case, the low energy effective Hamiltonian describing
interaction between neutrinos and matter is modified as follows:
\begin{eqnarray}
  \label{eq:neweffH}
  H_{\rm eff} = \sqrt{2} G_F N_e \left( \begin{array}{ccc}
    1 +  \varepsilon_{ee}     & \varepsilon_{e\mu} & \varepsilon_{e\tau} \\
      \varepsilon_{\mu e}  & \varepsilon_{\mu\mu}  & \varepsilon_{\mu\tau} \\
      \varepsilon_{\tau e} & \varepsilon_{\tau\mu} & \varepsilon_{\tau\tau}
               \end{array}
               \right)
\label{NSI-hamiltonian}
\end{eqnarray}
where $\epsilon$'s parameterize the nonstandard interactions (NSI)
of neutrinos with matter. Here, $G_F$ is the Fermi constant, $N_e$
denotes the averaged electron number density along the neutrino
trajectory in the earth. 

In this work we truncate the system so that we confine into the
$\mu - \tau$ sector of the neutrino evolution, which is justified
when $\epsilon$'s are sufficiently small. Then, the time evolution
of the neutrinos in flavor basis can be written as
\begin{eqnarray}
  \label{eq:evol}
i {d\over dt} \left( \begin{array}{c}
                   \nu_\mu \\ \nu_\tau
                   \end{array}  \right)
 = \left[ U \left( \begin{array}{cc}
                   0  & 0  \\
                   0   & \frac{ \Delta m^2_{32} }{2 E}
                   \end{array} \right)
            U^{\dagger} +  a \left( \begin{array}{cc}
            0 & \varepsilon_{\mu\tau} \\
             \varepsilon_{\tau\mu} & \varepsilon_{\tau\tau} -
                \varepsilon_{\mu\mu}
                   \end{array}
                   \right) \right] ~
\left( \begin{array}{c}
                    \nu_\mu \\ \nu_\tau
                   \end{array}  \right),
\end{eqnarray}
where $U$ is the flavor mixing matrix and $a \equiv \sqrt{2} G_F N_e$.
In the 2-2 element of the NSI term in the Hamiltonian is of the form
$\varepsilon_{\tau \tau} - \varepsilon_{\mu\mu}$ because the oscillation
probability depend upon $\varepsilon$'s only through this combination.
In the following, we set $\varepsilon_{\mu\mu} = 0$ for simplicity, and
study the sensitivity on $\varepsilon_{\tau\tau}$. At the end, the result
should be interpreted as $\varepsilon_{\tau \tau} - \varepsilon_{\mu\mu}$.
The evolution equation for the anti-neutrinos are given by changing the signs
of $a$ and replacing $U$ by $U^*$.

Since we work within the truncated 2 by 2 subsystem, we quote here
only the existing bounds of NSI parameters which are obtained
under the same approximation. By analyzing the Super-Kamiokande
atmospheric neutrino data the authors of \cite{constraint_nsi_atm}
obtained
\begin{equation}
| \varepsilon_{\mu\tau} | \lesssim \ 0.15, \ \ \ |
\varepsilon_{\tau\tau} | \lesssim \ 0.5,
\label{current_nsi_bounds}
\end{equation}
at 99 \% CL for 2 degrees of freedom.\footnote{
A less severe bound on $| \varepsilon_{\tau\tau} |$ is derived in
\cite{constraint_nsi_atm2} by analyzing the same data but with
$\varepsilon_{e\tau}$ and without $\varepsilon_{\mu \tau}$ }

\subsection{Numerical Results and Discussions}

\begin{figure}[bhtp]
\begin{center}
\includegraphics[height=18cm]{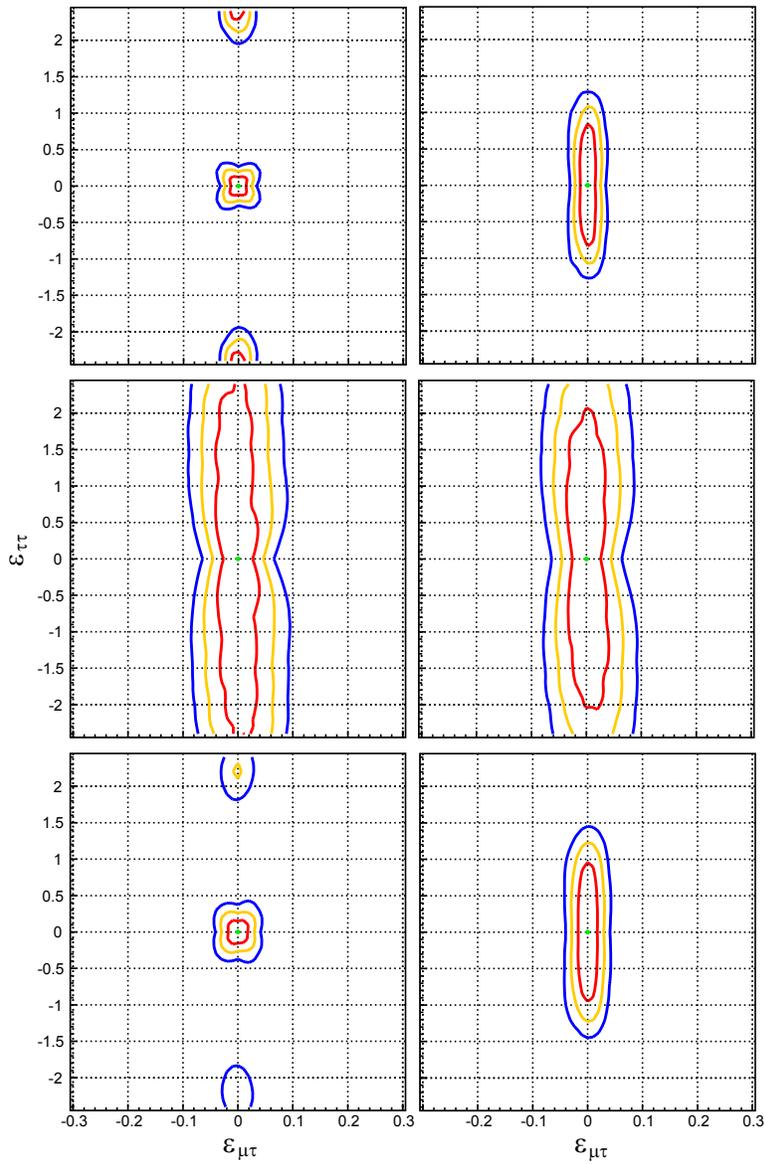}
\end{center}
\caption{
The allowed regions in
$\varepsilon_{\mu\tau} - \varepsilon_{\tau\tau}$ space for 4 years neutrino
and 4 years anti-neutrino running.
The upper, the middle, and the bottom three panels are for
the Kamioka-Korea setting, the Kamioka-only setting, and the
Korea-only setting,  respectively.
The left and the right panels are for cases with
$\sin^2 \theta \equiv \sin^2 \theta_{23} = 0.45$ and 0.5, respectively.
The red, the yellow, and the blue lines indicate the allowed regions
at 1$\sigma$, 2$\sigma$, and 3$\sigma$ CL, respectively.
The other input values of the parameters are idential to those
in Fig.~\ref{decoh-gamma-1-over-E}
}
\label{sens-kam_kor}
\end{figure}

In Fig.~\ref{sens-kam_kor}, presented are the allowed regions in
$\varepsilon_{\mu\tau} - \varepsilon_{\tau \tau}$
space for 4 years neutrino and 4 years anti-neutrino running of the
Kamioka-only  (upper panels),  the Korea-only (middle panels),
and the Kamioka-Korea (bottom panels) settings.
The input values $\varepsilon_{\mu\tau}$ and $\varepsilon_{\tau\tau}$
are taken to be vanishing.

As in the CPT-Lorentz violating case and unlike the system with
decoherence, the Korea-only setting gives much worse sensitivity
compared to the other two settings. Again the Kamioka-only setting
has a slightly better sensitivity than the Kamioka-Korea setting.
However we notice that the Kamioka-only setting has multiple
 $\varepsilon_{\tau\tau}$ solutions for $\sin^2 \theta_{23} = 0.45$.
The fake solutions are nearly eliminated in the Kamioka-Korea setting.

The sensitivities of three experimental setups at 2 $\sigma$ CL
can be read off from
Fig.~\ref{sens-kam_kor}. The approximate
2 $\sigma$ CL sensitivities of the
Kamioka-Korea setup  for
$\sin^2 \theta = 0.45 ~(\sin^2 \theta = 0.5)$ are:
\begin{equation}
| \epsilon_{\mu\tau} |  \lesssim  0.03 ~(0.03),~~~
| \epsilon_{\tau\tau}|  \lesssim  0.3 ~(1.2).
\end{equation}
Here we neglected a barely allowed region near $\epsilon_{\tau\tau} = 2.3$.
The Kamioka-only or Kamioka-Korea setup can improve the current
bounds on $\varepsilon$'s by factors of 8 (8) and 60 (16), which are
significant improvement.

There are a large number of references which studied the effects
of NSI and the sensitivity reach to NSI by the ongoing and the
various future projects. We quote here only the most recent ones
which focused on sensitivities by superbeam and reactor
experiments  \cite{KLOS} and neutrino factory \cite{CMNUZ}, from
which earlier references can be traced back.

By combining future superbeam experiment, T2K~\cite{T2K} and
reactor one, Double-Chooz \cite{reactor13}, the authors of
\cite{KLOS} obtained the sensitivity of $|\epsilon_{\mu \tau}|$ to
be $\sim$ 0.25 when it is assumed to be real (no CP phase) while
essentially no sensitivity to $\epsilon_{\tau \tau}$ is expected.
The same authors also consider the case of NO$\nu$A
experiment~\cite{NOVA} combined with some future upgraded reactor
experiment with larger detector as considered, e.g., in
\cite{DC200,angra} and obtained $\epsilon_{\mu \tau}$ sensitivity
of about 0.05 which is comparable to what we obtained.

While essentially no sensitivity of $\epsilon_{\tau \tau}$ is
expected by superbeam, future neutrino factory with the so called
golden channel $\nu_e \to \nu_\mu$ and $\bar{\nu}_e \to
\bar{\nu}_\mu$, could reach the sensitivity to $\epsilon_{\tau
\tau}$ at the level of $\sim$ 0.1-0.2~\cite{CMNUZ}. Despite that
the sensitivity to $\epsilon_{\mu \tau}$ by neutrino factory was
not derived in ~\cite{CMNUZ}, from Fig. 1 of this reference, one
can naively expect that the sensitivity to $\epsilon_{\mu \tau}$
is similar to that of $\epsilon_{ee}$ which is  $\sim$ 0.1 or so.
We conclude that the sensitivity we obtained for $\epsilon_{\mu
\tau}$ is not bad.

\section{Conclusion \label{conclusion}}

The Kamioka-Korea two detector system for the J-PARC neutrino beam
was shown to be a powerful experimental setup for lifting the
neutrino parameter degenracies and probing CP violation in
neutrino oscillations. In this talk, I presented the sensitivities
of the same setup to nonstandard neutrino physics such as quantum
decoherence, tiny violation of Lorentz symmetry, and nonstandard
neutrino interactions with matter. Generally speaking, two
detector setup is more powerful than one detector setup at
Kamioka, not only for lifting the neutrino parameter degeneracies,
but also probing/constraining nonstandard neutrino physics. The
sensitivities of three experimental setups at 90\% CL are
summarized in Table \ref{table-gamma} and Table \ref{table-liv}
for quantum decoherence and Lorentz symmetry violation
with/without CPT symmetry, respectively. We can say modestly that
future long baseline experiments with two detector setup can
improve the sensitivities on nonstandard neutrino physics in many
cases, in addition to lifting the neutrino parameter degeneracies.
It would be highly desirable to make such neutrino physics
facilities realitic in the near future.

\section*{}
PK is grateful to the organizers of the workshops for inviting him for the talk.  




\end{document}